\documentclass[sigconf, nonacm]{acmart}
\AtBeginDocument{%
  }

\setcopyright{acmlicensed}
\copyrightyear{2018}
\acmYear{2018}
\acmDOI{XXXXXXX.XXXXXXX}
\acmConference[Conference acronym 'XX]{Make sure to enter the correct
  conference title from your rights confirmation email}{June 03--05,
  2018}{Woodstock, NY}
\acmISBN{978-1-4503-XXXX-X/2018/06}

\usepackage{algorithm}
\usepackage{algpseudocode}
\algrenewcommand\algorithmicensure{\textbf{Return:}}

\usepackage{multirow}
\usepackage{booktabs}
\usepackage{textcomp}

\begin{document}

\title{Getting a GRIP on Graph-Based Search \\ for Learned Sparse Retrieval}

\author{Vihan Lakshman}
\email{vihan@mit.edu}
\affiliation{%
  \institution{MIT CSAIL}
  \city{Cambridge}
  \country{USA}
}

\author{Franco Maria Nardini}
\email{francomaria.nardini@isti.cnr.it}
\affiliation{%
  \institution{ISTI–CNR}
  \city{Pisa}
  \country{Italy}
}

\author{Cosimo Rulli}
\email{cosimo@linkup.so}
\affiliation{%
  \institution{Linkup}
  \city{Paris}
  \country{France}
}

\author{Rossano Venturini}
\email{rossano.venturini@unipi.it}
\affiliation{%
  \institution{University of Pisa}
  \city{Pisa}
  \country{Italy}
}

\renewcommand{\shortauthors}{Lakshman et al.}

\begin{abstract}
Learned sparse retrieval (LSR) has recently gained traction in the information retrieval community for the ability to combine the generalization capabilities of neural representations with the interpretability of keyword-based search. A number of recent works have demonstrated that inverted indexes, when appropriately modified for learned sparse embeddings, achieve state-of-the-art recall-latency tradeoffs for LSR, often outperforming graph-based indexes by substantial margins in latency at a fixed recall target. As a result, the community has generally concluded that graph-based approximate nearest neighbor (ANN) search is inferior to inverted indexes for LSR. This paper challenges this conventional wisdom. Our key insight is that, while applying standard graph-based ANN indexes out-of-the-box to sparse vectors leads to uncompetitive performance, we can overcome these shortcomings by infusing the strengths of inverted indexes \emph{into} graph-based search. In particular, we introduce three techniques to accelerate graph-based LSR: (i) \emph{term-based entry point seeding}, (ii) \emph{approximate routing via sparse sketching}, and (iii) \emph{sparse partition-based graph construction}. We combine these techniques into a new graph-based ANN system for LSR we call \textsc{GRIP}. On the MS MARCO v2 benchmark (138M passages) with SPLADE embeddings, \textsc{GRIP} achieves a $2.5\times$ latency improvement over Seismic, a state-of-the-art inverted index tailored for LSR, at 95\% recall@10. Ultimately, our results call into question the default choice of inverted indexes for deploying learned sparse retrieval.
\end{abstract}

\maketitle

\section{Introduction}
Learned sparse embeddings have recently gained popularity in information retrieval applications by bridging the strengths of dense neural representations and interpretable lexical search. Consequently, a number of works have studied building indexes to enable efficient sparse similarity search at scale. Following the breakthrough success of systems such as Seismic~\cite{bruch2024efficient}, inverted index designs currently constitute the state of the art for sparse vector search, substantially outperforming graph-based methods, which typically achieve the strongest performance on \emph{dense} retrieval workloads~\cite{aumuller2020ann}. 

The intuition behind the success of inverted indexes for learned sparse retrieval is simple and compelling: sparse neural representations maintain many structural similarities to classical bag-of-words and TF-IDF sparse representations of text, and inverted indexes were precisely designed for efficient search over such structures. Graph-based indexes, on the other hand, were conceived to address dense vector retrieval, a fundamentally different problem, and graphs are generally considered ill-suited for learned sparse retrieval due to network traversals incurring many redundant distance computations and poor memory access patterns~\cite{li2026sindi}. 

\begin{figure}[h]
  \centering
  \includegraphics[width=0.8\linewidth]{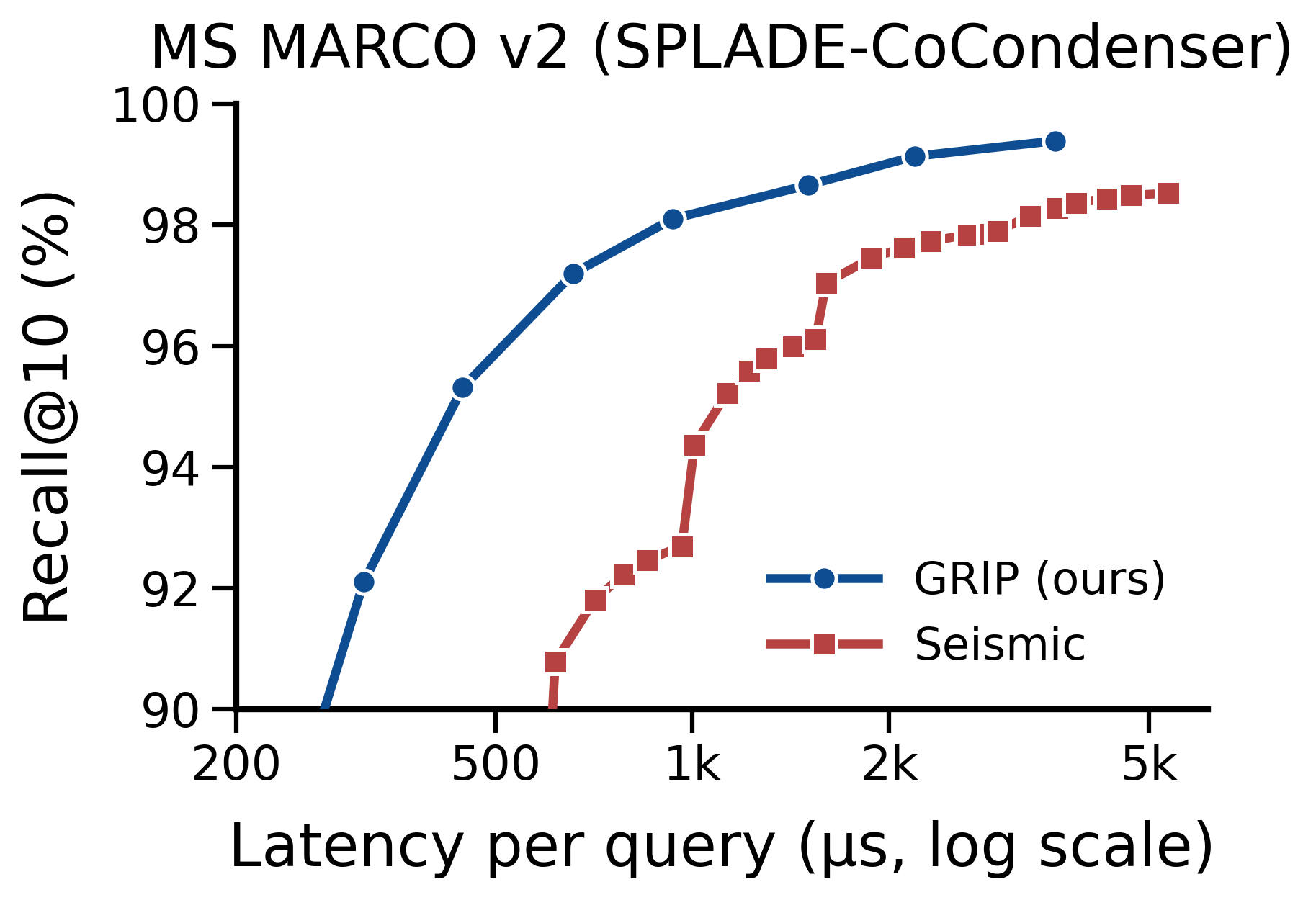}
  \caption{On the MS MARCO v2 search benchmark with SPLADE-CoCondenser embeddings, \textsc{GRIP} outperforms Seismic, a state-of-the-art inverted index system for LSR, by $2.2-3.4\times$ in latency across all fixed recall targets.}
  \label{fig:grip-v2}
\end{figure}

This paper challenges this conventional wisdom that graph-based methods are inherently inferior to inverted indexes for learned sparse retrieval. To that end, we present \textsc{GRIP} (\textbf{G}raph \textbf{R}etrieval for Sparse \textbf{I}nner \textbf{P}roduct Search), a graph-based approximate nearest-neighbor search system for sparse vector search that achieves a $2.2$--$3.4\times$ latency improvement over Seismic at recall@10 targets between 90\% and 98\% on the full MS MARCO v2 benchmark (138M passages) with SPLADE-CoCondenser sparse embeddings, including a $2.5\times$ improvement at 95\% recall@10. On MS MARCO v1 (8.84M passages), \textsc{GRIP} is $1.4$--$1.7\times$ faster than Seismic with SPLADE-CoCondenser embeddings at every recall target from 90\% to 99\%. With SPARSEUP~\cite{Formal_2026}, a more recent sparse embedding model, \textsc{GRIP} is $1.2\times$ faster than Seismic at 92--95\% recall@10 and on par at 97\%, while Seismic retains an edge in the extremely high-recall regime (98--99\%). 

Our core insight driving the design of \textsc{GRIP} is that we can infuse the strengths of inverted indexes into graph-based search such that we retain the benefits of both index types. First and foremost, the signature feature of inverted indexes is that a sparse query vector only needs to touch the posting lists associated with its non-zero vector components. For sufficiently sparse vectors, this drastically reduces the search space. Standard graph-based indexes discard this structure. \textsc{GRIP} explicitly ports this structural advantage of the inverted index into graph search via a technique we call \emph{term-based entry point seeding}, which precomputes, for every vocabulary term, the documents with the largest weight on that term, and initializes the graph search from the documents listed under the query's highest-impact terms. Better entry point selection for graph search is a well-studied problem in the vector database literature, both for dense retrieval~\cite{oguri2024theoretical} and by seeding graph traversal with the results of a separate lexical retrieval stage~\cite{kulkarni2023lexically}. \textsc{GRIP} differs from prior work in using the sparse embeddings themselves as the entry point selection mechanism, which requires no additional retrieval system and almost no additional work at query time, making graph search $1.95-2.59\times$ faster on its own in our experimental evaluations.

Secondly, we further accelerate \textsc{GRIP} via an optimization we call \emph{approximate routing via sparse sketching} that is also inspired by inverted index designs for LSR. Learned sparse embeddings exhibit \emph{concentration of importance}~\cite{bruch2024efficient}: a small number of components carry most of each vector's weight. We exploit this property by sketching each document as its $R$ largest-magnitude components and running the graph traversal entirely on these sketches, then reranking the final candidates against the full vectors. Because learned sparse weights are non-negative, a sketched score is always a lower bound on the exact score, so only the navigation of the graph is approximate, and the returned results are scored exactly. This sketching is complementary to quantizing the stored weights, which we also apply; sketching reduces how many components each distance computation touches, while quantization reduces the cost of each one.

Our final key optimization is not related to accelerating the query time of \textsc{GRIP} but rather the graph index construction. Traditionally, graph-based vector search indexes are constructed by calling the query function as a subroutine, inserting each new node by searching for its nearest neighbors among the documents already in the graph and adding edges according to a set of heuristics. As we show in this paper, sparse vectors are ill-suited to this search-based construction approach: while graphs over dense embeddings typically see diminishing returns beyond a construction beam width (ef\_construction) of roughly 200~\cite{elliott2024impacts}, we observe that HNSW graphs over learned sparse embeddings require a beam width of 2000 to achieve competitive search quality (when combined with our other aforementioned optimizations). However, this increase blows up the graph construction time by roughly an order of magnitude, and on the MS MARCO v2 dataset, we estimate that we would require several days to build the graph. To solve this problem, we take inspiration from PiPNN~\cite{rubel2026pipnn}, a recently proposed index construction  algorithm that replaces the incremental search-based procedure of prior systems with vector partitioning and hashing. PiPNN's partitioning step, which assigns points to randomly sampled leaders, closely mirrors the shallow $k$-means clustering that Seismic uses to organize its posting lists, and we find that it partitions learned sparse vectors effectively. 

However, we cannot apply PiPNN to learned sparse vectors out-of-the-box because PiPNN computes nearest neighbors within each partition using dense matrix multiplication, which is wasteful for sparse vectors spanning tens of thousands of  dimensions. Again taking inspiration from inverted index designs for LSR, we instead compute these inner products from a small inverted index built over each partition. We apply this approach to develop a sparse variant of PiPNN we call \emph{sparse partition-based graph construction}, followed by a single refinement pass in which each document searches the finished graph from its own term seeds. This reduces the time to construct a \textsc{GRIP} graph from 5.5 hours (with HNSW) to 47 minutes on the 8.8M MS MARCO v1 collection with SPLADE-CoCondenser embeddings. While we show that \textsc{GRIP}'s  sparse PiPNN graph is somewhat weaker than an HNSW graph under standard search, the order-of-magnitude improvement in construction time allows us to build the graph for MS MARCO v2 in hours (7.3 hours) as opposed to days.

In addition to improving our scientific understanding of index design for learned sparse retrieval, enabling graph-based indexes to be more efficient for sparse vector workloads also has important practical implications. Many modern retrieval systems now deploy so-called \emph{hybrid} search strategies where dense and sparse retrieval are combined to produce a final list of results. Since such systems already typically deploy graph-based indexes for dense retrieval, they may be able to consolidate and simplify their infrastructure by reusing the same components for sparse search as opposed to having to maintain separate inverted index logic. Moreover, the popularity of graph-based indexes has led to a mature ecosystem of high-performance libraries and remains an active area of research. Thus, efficient graph-based search for LSR can enable the sparse retrieval community to also take advantage of these ongoing developments.

In summary, this paper makes the following contributions:

\begin{itemize}
\item \textbf{End-to-End System:} We challenge the conventional wisdom that graph-based indexes for sparse vector similarity search are uncompetitive with inverted indexes at practically relevant document collection sizes. We support this claim by presenting \textsc{GRIP}, a new system for graph-based learned sparse retrieval that achieves substantial performance improvements over standard graph-based search designed for dense retrieval via three algorithmic techniques which we highlight further below. These results call into question the default choice of inverted indexes for deploying learned sparse retrieval.

\item \textbf{Term-based Entry Point Seeding:} We present a new approach for selecting entry points for graph-based similarity search tailored to the structure of learned sparse embeddings. In particular, we precompute a list of documents with the largest weights per vocabulary term and initialize graph search at the nodes associated with the query's highest-impact components.  

\item \textbf{Approximate Routing:} Inspired by the concentration of importance property of learned sparse embeddings, we propose a simple and effective sketching technique that retains a document embedding's $R$ largest-magnitude components and performs graph traversal entirely on the sketched vectors. We observe that this reduces query latency with negligible loss in search quality. 

\item \textbf{Sparse PiPNN:} We find that the traditional search-based iterative graph construction procedures used by popular indexes such as HNSW and DiskANN are especially poor fits for LSR because sparse embeddings require searching a much larger pool of candidates to build sufficiently navigable graphs. While applying traditional graph-based construction methods, in conjunction with our aforementioned optimizations, can produce competitive performance with inverted indexes, the build time is prohibitively expensive at scale. To remedy this bottleneck, we take inspiration from the recently proposed PiPNN construction algorithm~\cite{rubel2026pipnn} and adapt PiPNN to the sparse setting by utilizing inverted indexes in the construction process. 

\end{itemize}

\section{Related Work}

\subsection{Learned Sparse Embeddings}

Sparse vector search is a foundational problem in information retrieval dating back to early foundational work on bag-of-words and term frequency representations of text~\cite{schutze2008introduction}. However, traditional sparse lexical vectors are fundamentally limited in that they can only match queries to documents with exact keyword overlaps. \emph{Learned} sparse representations were conceived to address this shortcoming by utilizing neural networks to learn high-dimensional representations of vectors with a small number of nonzero components such that each component represents a generalized concept as opposed to a fixed word. Following the revolution in transformer-based representation learning, a number of works have studied how to train or fine tune learned sparse representations~\cite{bai2020sparterm, dai2019context, dai2020context, dai2020context-weighting, macavaney2020expansion, lin2021few, zhao2021sparta}.

One of the most influential families of learned sparse embeddings models is SPLADE~\cite{formal2021splade, formal2021splade-v2, lassance2022efficiency, lassance2024splade} which established learned sparse retrieval as a state-of-the-art technique for first-stage text retrieval. Recently, \citet{Formal_2026} introduced SPARSEUP, a next-generation learned sparse embedding model that achieves improved retrieval quality over SPLADE and, to our knowledge, currently constitutes the best-in-class sparse embedding model for first-stage retrieval. Furthermore, another very recent line of work has studied the notion of \emph{inference-free} learned sparse embeddings which replace the query encoder model with a learned lookup table~\cite{nardini2025effective, shen2025exploring}. 

\subsection{Algorithms for Learned Sparse Retrieval}

In addition to training models for encoding effective learned sparse embeddings, the retrieval community has also proposed numerous techniques for efficiently searching over these representations. A key finding in early work within this area is that learned sparse representations generate term distributions that differ considerably from the Zipfian nature of bag-of-words text representations, rendering traditional inverted indexes designed for text matching as inefficient~\cite{mackenzie2021wacky, bruch2023approximate}. 

Several of the early specialized indexes for learned sparse retrieval involved graph-based search. The Neurips 2023 Big ANN benchmarking competition~\cite{simhadri2024results} included a dedicated learned sparse retrieval track where the winning solutions, PyANNS and GrassRMA, utilized graph-based vector search, a dominant approach for dense vector retrieval~\cite{aumuller2020ann}, for learned sparse similarity search. Both PyANNS and GrassRMA utilize the HNSW graph index~\cite{malkov2018efficient} with additional low-level optimizations, such as integer quantization of values and graph reordering~\cite{coleman2022graph}, to improve performance. Notably, these methods do not seek to infuse the structural properties of inverted indexes into graph-based search, which, at a methodological level, differentiates \textsc{GRIP}. 

These graph-based methods subsequently lost their standing as the state-of-the-art indexing solution for LSR following the introduction of Seismic~\cite{bruch2024efficient} which achieved substantial efficiency improvements over PyANNS and GrassRMA using an inverted index tailored to the idiosyncrasies of learned sparse embeddings. Seismic has since been further improved in a number of follow-on works~\cite{bruch2025investigating, bruch2024pairing, nardini2026compact}. In addition to Seismic, several other systems have proposed efficient inverted indexes for LSR including Block-Max Pruning (BMP)~\cite{mallia2024faster}, Dynamic Superblock Pruning (DSP)~\cite{carlson2025dynamic}, and SINDI~\cite{li2026sindi} that each seek to accelerate the posting list traversal component of inverted index-based search. The latest version of Seismic, described in the work of ~\cite{nardini2026compact}, includes further performance optimizations that achieve significant search quality-efficiency improvements over both other inverted index designs for LSR and graph-based approaches and, to our knowledge, represents the current state-of-the-art system for LSR. Thus, we treat this latest version of Seismic as our primary baseline when comparing with our proposed \textsc{GRIP} system. 

\section{Methodology}

We now present the algorithmic details behind \textsc{GRIP}. As in prior work on graph-based search indexes, \textsc{GRIP} consists of an \emph{offline} index build phase and an \emph{online} search phase. Algorithm~\ref{alg:seed-lists} and Algorithm~\ref{alg:sparse-pipnn} together describe the offline phase where we build a data structure of term-seeded entry points and construct the graph for traversal via our sparse PiPNN algorithm, respectively. Algorithm~\ref{alg:grip-search} then describes the online search procedure which follows the standard paradigm of graph-based search for ANN via beam search with two modifications: (i) we initialize the graph search with up to $T \cdot L$ nodes, the $L$ highest-weighted documents (from the precomputed term seed lists) for each of the $T$ highest-impact query terms, and (ii) we perform the traversal using top-$R$ sketched sparse vectors and re-score the final top-$k$ results with the full embeddings.

\begin{algorithm}[t]
  \caption{Building the term seed lists}
  \label{alg:seed-lists}
  \begin{algorithmic}[1]
  \Require Collection $\mathcal{D} = \{x_1, \dots, x_n\}$ of sparse vectors over vocabulary $V$; seeds per term $L$
  \Ensure Seed lists $\{H_t\}_{t \in V}$; per-term maximum weights $\{w_t\}_{t \in V}$
  \For{each term $t \in V$}
    \State $H_t \gets$ empty min-heap of capacity $L$, keyed on weight
    \State $w_t \gets 0$
  \EndFor
  \For{each document $d \in \mathcal{D}$}
    \For{each term $t$ with $x_d[t] > 0$}
      \State $w_t \gets \max(w_t, x_d[t])$
      \If{$|H_t| < L$}
        \State push $(x_d[t], d)$ onto $H_t$
      \ElsIf{$x_d[t] > \min(H_t)$}
        \State replace the minimum of $H_t$ with $(x_d[t], d)$
      \EndIf
    \EndFor
  \EndFor
  \State \Return $\{H_t\}$, $\{w_t\}$
  \end{algorithmic}
  \end{algorithm}

\begin{algorithm}[t]
  \caption{Sparse PiPNN}
  \label{alg:sparse-pipnn}
  \begin{algorithmic}[1]
  \Require Sparse vectors $\{x_1, \dots, x_n\}$; neighbors per point within a leaf $k_\ell$; degree bound $D$
  \Ensure Graph $G$ with out-degree at most $D$
  \State $\mathcal{P} \gets \Call{BallCarve}{\{x_i\}}$ \Comment{\textbf{Sparse:} leaders scored via inverted index. See \cite{rubel2026pipnn} for the full ball carving description.}
  \For{each leaf $P \in \mathcal{P}$}
    \State build an inverted index over $P$ \Comment{\textbf{Sparse:} replaces PiPNN's dense GEMM}
    \State compute all pairwise inner products in $P$ from its posting lists
    \State add each point's $k_\ell$ nearest neighbors in $P$ as candidate edges
  \EndFor
  \State $\mathcal{C} \gets \Call{HashPrune}{\text{candidates}}$ \Comment{\citet{rubel2026pipnn}}
  \State $G \gets \Call{RobustPrune}{\mathcal{C}, D}$ \Comment{PiPNN, on inner products}
  \For{each point $p$} \Comment{\textbf{Sparse:} refinement pass}
    \State $\textit{cand} \gets$ beam search of $G$ for $x_p$ from $p$'s term seeds
    \State $N(p) \gets \Call{RobustPrune}{\textit{cand} \cup N(p), D}$
  \EndFor
  \end{algorithmic}
  \end{algorithm}

\begin{algorithm}[t]
  \caption{\textsc{GRIP} Search}
  \label{alg:grip-search}
  \begin{algorithmic}[1]
  \Require Query $q$; graph $G$; full vectors $\{x_d\}$; 8-bit routing sketches $\{\tilde{x}_d\}$ of each document's top-$R$ weights; seed
  lists $\{H_t\}$ and maxima $\{w_t\}$; seed terms $T$; beam width $\textit{ef}$; patience $P$
  \Ensure Top-$k$ documents for $q$
  \State $\tilde{q} \gets$ 8-bit quantization of $q$ \Comment{integer query table}
  \State $\mathcal{T} \gets$ the $T$ query terms of largest $q_t \cdot w_t$
  \State $S \gets \bigcup_{t \in \mathcal{T}} H_t$ \Comment{term-based entry points}
  \State $B \gets \Call{BeamSearch}{G, \tilde{q}, \{\tilde{x}_d\}, S, \textit{ef}, P}$ \Comment{routing on sketches; stop after $P$ expansions without top-$k$
  change}
  \State \Return top $k$ of $B$ rescored with $\langle q, x_d \rangle$ \Comment{exact rescoring}
  \end{algorithmic}
  \end{algorithm}

\subsection{Term-Seeded Entry Point Selection}

While prior work in vector similarity search has studied judiciously choosing entry points for graph-based search~\cite{oguri2024theoretical, kulkarni2023lexically}, no research, to our knowledge, has examined how to choose entry points for learned sparse embeddings or has demonstrated that graph-based search for LSR benefits from entry point selection. We close this gap. Our core insight is that standard graph-based search begins every query from a fixed, query-independent entry point, and must then traverse the graph to converge to the neighborhood of the query's nearest neighbors, paying for distance computations at every hop along the way. An inverted index, in contrast, needs no such traversal: the query's non-zero components directly identify the posting lists where its answers are guaranteed to reside. Term-based entry point selection brings this advantage to graph search by initializing the search at nodes that carry large weight on the query's most important terms, which leads to fewer wasted distance computations. Specifically, in the offline phase, described in Algorithm~\ref{alg:seed-lists}, we precompute for every vocabulary term $t \in V$ a seed list $H_t$ containing the $L$ documents with the largest weight on that term, that is, the $L$ documents $d$ maximizing $x_d[t]$, along with the largest weight on the term, $w_t = \max_{d} x_d[t]$. Each seed list is maintained as a min-heap of capacity $L$, so the lists can be built in a single pass over the non-zero entries of the collection in $O(\mathrm{nnz} \cdot \log L)$ time. The resulting table holds at most $|V| \cdot L$ entries which concretely translates to roughly 8 MB of space on the MS MARCO v1 8.8M passage collection, a negligible space overhead in comparison to the 8.8 GB used to store the remainder of the index. At query time, we use $w_t$ to bound how much each query term can contribute to any document's score: for every document $d$, $q_t \cdot x_d[t] \le q_t \cdot w_t$. We select the $T$ query terms with the largest such upper bound, take the union of their seed lists $S = \bigcup_{t \in \mathcal{T}} H_t$, which contains at most $T \cdot L$ documents, and initialize the beam with every document in $S$. We set $L = 64$ throughout and $T = 2$ for SPLADE-CoCondenser and $T = 4$ for SPARSEUP, each selected on held-out training queries disjoint from the evaluation queries.

\subsection{Approximate Routing via Sparse Sketching}

The second core optimization at the heart of \textsc{GRIP} is approximate routing via sparse sketching, which is inspired by the concentration of importance phenomenon of learned sparse embeddings identified by \citet{bruch2024efficient}. In essence, concentration of importance states that a small fraction of the non-zero components of learned sparse vectors, which themselves are a small proportion of the total number of dimensions, contains the majority of the $\ell_1$ mass of the vector. While \citet{bruch2024efficient} utilize this insight in the design of Seismic, we leverage this same structure for faster graph-based traversal. To our knowledge, this is a novel optimization in the context of sparse graph ANN since prior graph-based systems for LSR such as PyANNS and GrassRMA adopt quantization techniques to represent the vectors in lower precision (as we also do in \textsc{GRIP}) but do not consider dropping low-magnitude non-zero components entirely. In particular, our sparse sketching scheme, as described in Algorithm~\ref{alg:grip-search}, involves constructing, for every document $d$, a sketch $\tilde{x}_d$ that retains only the $R$ largest-magnitude non-zero components of $x_d$ and sets the remaining components to zero. Importantly, sketching affects only search: the graph itself is constructed on the full vectors, as are the seed lists, so the edges of the graph reflect exact similarities between documents. 

We store these sketches as a second copy of the collection, quantized to 8 bits, and run the entire beam search over them: both the choice of which nodes to expand and the decision of when to terminate are made using the sketched scores $\langle q, \tilde{x}_d \rangle$. Once the search terminates, we rescore every document in the final beam with its exact score $\langle q, x_d \rangle$ against the full vectors and return the top $k$. Because the weights of learned sparse embeddings are non-negative, the sketched score is always a lower bound on the exact score, $\langle q, \tilde{x}_d \rangle \le \langle q, x_d \rangle$, and the error is confined to the documents' smallest components. Thus, the sketch only perturbs the route the search takes through the graph, while the returned results are always ranked by their exact scores. Moreover, when $R$ is at least the number of non-zero components of every document, sketched search reduces to search over the full (quantized) vectors. On SPLADE-CoCondenser, whose documents have 127 non-zero components on average, we set $R = 80$, which keeps only 62\% of the non-zero entries while speeding up search by $1.22$--$1.38\times$ at recall@10 targets between 95\% and 99\%. We use $R = 128$ for SPARSEUP, whose documents have 190 non-zero components on average. For both experiments, we set the value of $R$ via holdout training queries. 

\subsection{Sparse PiPNN}

\begin{figure}[h]
  \centering
  \includegraphics[width=0.55\textwidth]{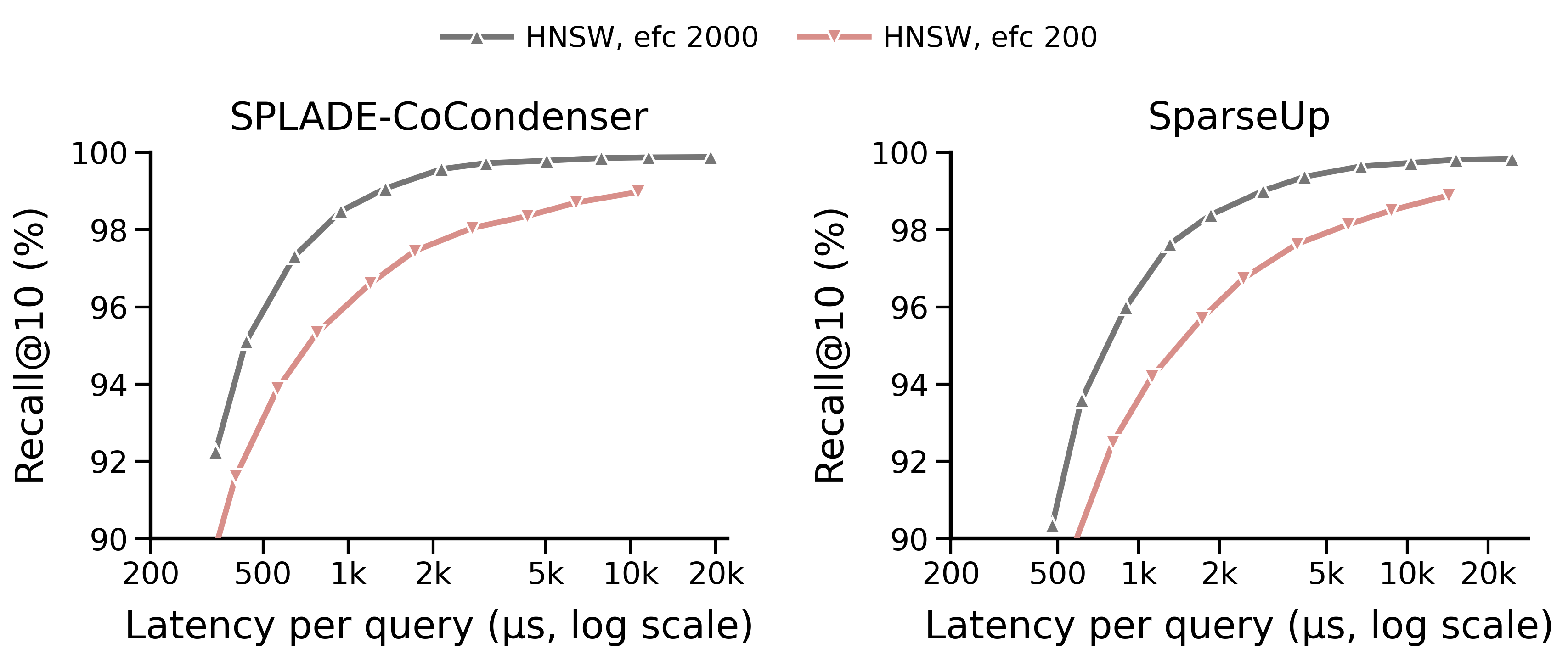}
  \caption{Unlike in dense vector retrieval, where graph-based systems like HNSW achieve diminishing returns with an ef construction parameter above 200, learned sparse workloads continue to improve in quality with an ef construction value up to 2000. }
  \label{fig:hnsw-efc}
\end{figure}

\begin{figure}[h]
  \centering
  \includegraphics[width=\linewidth]{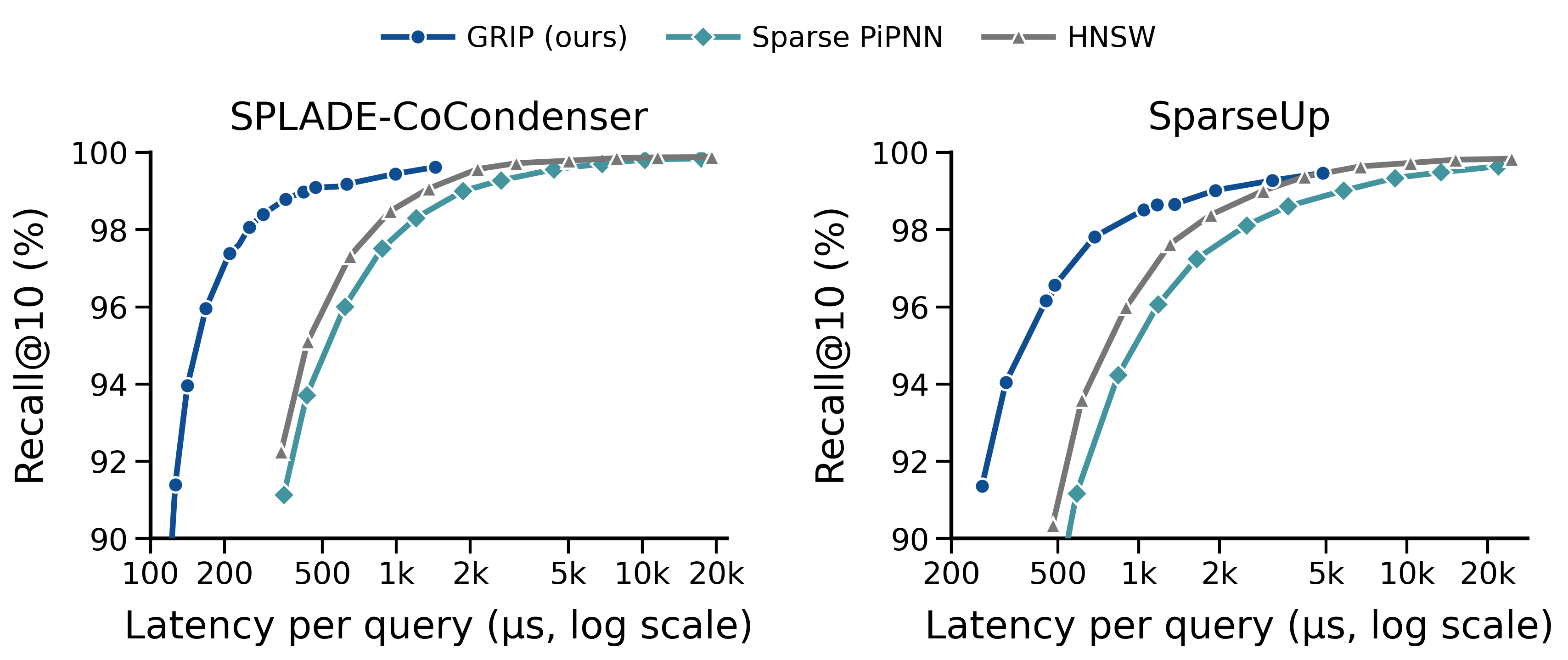}
  \caption{We observe that Sparse PiPNN achieves close to the performance of standard HNSW, but at a fraction of the index build time. Furthermore, \textsc{GRIP}, which augments the Sparse PiPNN graph with the term-seeding and approximate routing, outperforms HNSW.}
  \label{fig:grip-vanilla}
\end{figure}

We now discuss the third core optimization behind \textsc{GRIP}, namely sparse partitioned graph construction. To motivate the need for a new graph construction approach that differs from the traditional paradigm of building navigable graphs by iteratively inserting documents via repeated calls to the beam search subroutine, we demonstrate in Figure~\ref{fig:hnsw-efc} that on learned sparse embeddings the quality of graph search improves substantially when the construction beam width (efc) is increased from 200 to 2000, while dense retrieval typically sees diminishing returns on search quality with efc values in the range of 100 to 300~\cite{elliott2024impacts, malkov2018efficient}. On MS MARCO v1, an HNSW graph built with an efc of 2000 is $1.7$--$3.5\times$ faster to search than one built with an efc of 200 at recall@10 targets between 95\% and 98\%, and the latter cannot exceed roughly 99\% recall@10 at all. This finding further underscores the widely held observation that graph-based search is not particularly well-suited to sparse retrieval without further innovation. While we can overcome this shortcoming simply by increasing the efc parameter, this solution is not scalable as it requires 5.5--7.4 hours to construct an HNSW graph on MS MARCO v1 and extrapolating from those numbers, several days to build such a graph on MS MARCO v2's 138M passages. This is impractical for collections that must be reindexed frequently.

To circumvent this prohibitively long index construction time, we turn to PiPNN~\cite{rubel2026pipnn}, a recent algorithmic breakthrough in efficiently constructing graph-based ANN indexes that completely avoids the iterative search-based insertion procedure. Unfortunately, we cannot use PiPNN directly out-of-the-box for LSR because several steps of the algorithm are specifically designed for dense vectors. As described in Algorithm~\ref{alg:sparse-pipnn}, we modify the original PiPNN construction procedure to be amenable to sparse vectors. 

The key insight behind our \emph{Sparse PiPNN} approach is to replace PiPNN's dense vector multiplication subroutines with sparse computations by utilizing inverted indexes to efficiently compute pairwise inner products. In particular, we apply this idea at the two stages of PiPNN that are dominated by dense computation. First, PiPNN partitions the collection by randomized ball carving: it samples a set of leaders, assigns every point to its nearest few leaders, and recurses until every partition holds at most $C$ points. Rather than comparing every point against every leader, we build an inverted index over the leaders' non-zero components, so each point accumulates scores only for the leaders with which it shares at least one term; a point that shares no term with any leader is assigned to a leader chosen by hashing, so that such points do not all collapse into the same partition. 

Second, within each leaf partition $P$, PiPNN computes all pairwise similarities with a dense matrix multiplication, at a cost of $O(|P|^2 \cdot |V|)$, which is wasteful when vectors span tens of thousands of dimensions but have only around a hundred non-zero components each. We instead build a small inverted index over the documents in $P$ and accumulate every pairwise inner product by iterating over its posting lists, at a cost of $\sum_{t} |P_t|^2 / 2$, where $P_t$ denotes the documents in $P$ with a non-zero weight on term $t$. The remaining steps of PiPNN, HashPrune and RobustPrune, carry over directly, with similarities measured by inner product rather than Euclidean distance throughout. Finally, deviating from the original PiPNN algorithm, we add a final refinement pass in which each document searches the finished graph starting from its own term-based seeds and re-prunes its neighbor list with the results. Although this pass performs one search per document, it searches a completed graph rather than inserting documents one at a time, and we find that it is important for search quality at very high recall targets. 

With leaves of at most 1{,}024 documents and a maximum out-degree of 64, Sparse PiPNN builds the graph for the 8.8M-passage MS MARCO v1 collection in 47 minutes, compared to 5.5 hours for HNSW with an efc of 2000. As with term-seeded entry point selection, where we borrow ideas from inverted indexes to initialize graph-based search, and approximate routing, where we take inspiration from the concentration of importance property used in the Seismic inverted index system, our sparse PiPNN algorithm utilizes inverted indexes as part of the index build stage to reduce graph construction time by an order of magnitude. In each of our core optimizations behind \textsc{GRIP}, we see a common narrative pattern of infusing the strengths of inverted index designs to make graph-based search for LSR more performant.

As shown in Figure~\ref{fig:grip-vanilla}, a graph built with our Sparse PiPNN approach (without any of the other \textsc{GRIP}-specific optimizations) performs slightly below standard HNSW search with an efc construction setting of 2000 while building in 47 minutes as opposed to 5.5 hours. Thus, Sparse PiPNN trades off a small loss in search quality for major gains in construction build time, which becomes particularly important at larger collection sizes. Furthermore, we also note in Figure~\ref{fig:grip-vanilla} that GRIP, which adds entry point seeding and approximate routing on top of Sparse PiPNN, outperforms Sparse PiPNN and HNSW, which underscores the value of our other aforementioned optimizations.

\section{Experiments}

We implemented \textsc{GRIP} in Rust, building off the ANN search primitives developed within the \textsc{kannolo} library~\cite{delfino2025kann}. We conduct our experimental evaluation using two standard sparse retrieval benchmarks, MS MARCO v1, which consists of 8.8M passages and MS MARCO v2, which contains 138M passages in the collection. We embed both sets of passages using two different sparse embedding models: SPLADE-CoCondenser~\cite{formal2022distillation} and the recently released SPARSEUP model~\cite{Formal_2026} which achieves state-of-the-art performance on several retrieval benchmarks. We conducted all experiments on a single cloud instance with an Intel Xeon Platinum 8358 @ 2.60 GHz processor, 30 vCPUs, and 222 GB of RAM. We use all 30 available threads for index construction and a single thread at query time. 

We compare \textsc{GRIP} against two representative baselines. The first is HNSW without any sparsity-specific optimizations to provide a reference point for \textsc{GRIP}'s specific improvements tailored to LSR. We use the \textsc{kannolo} Rust implementation of HNSW. Second, we also compare with the latest version of Seismic~\cite{bruch2024efficient}, using the official Rust implementation, which represents, to the best of our knowledge, the current state-of-the-art index for LSR in terms of recall-latency tradeoffs. We do not explicitly benchmark other LSR indexes such as SINDI~\cite{li2026sindi}, GrassRMA, and PyANNS since recent work has shown that an optimized implementation of Seismic outperforms these other methods by substantial margins~\cite{nardini2026compact}. 

\begin{figure}[h]
  \centering
  \includegraphics[width=1.1\linewidth]{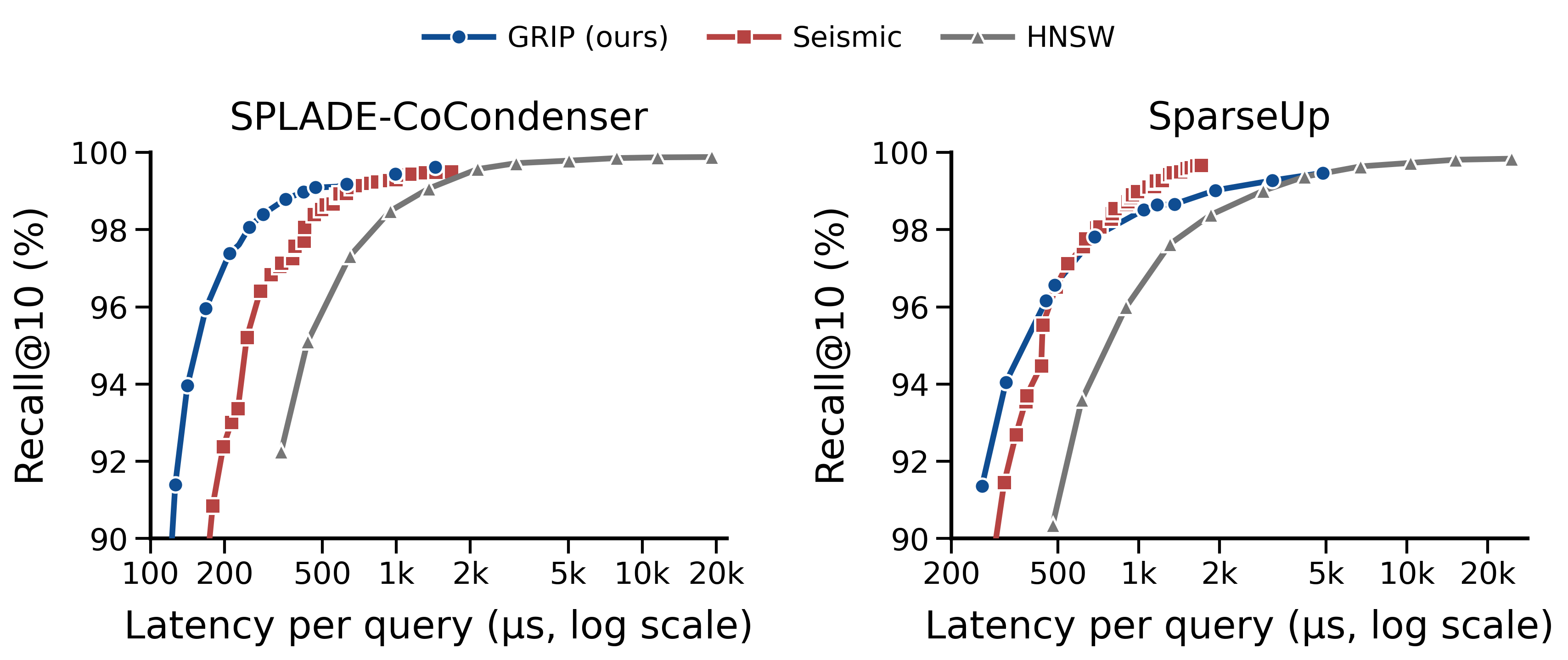}
  \caption{On the MS MARCO v1 dataset, we observe that \textsc{GRIP} is $1.4-1.7\times$ faster than Seismic across all recall@10 targets plotted with SPLADE-CoCondenser embeddings. With SPARSEUP embeddings, \textsc{GRIP} is $1.2\times$ faster than Seismic
  at 92–95\% recall@10 and roughly on par at 97\%, while Seismic remains faster in the extremely high-recall regime of 98\% and above.}
  \label{fig:grip-v1}
\end{figure}

We report our primary experimental findings in Figures~\ref{fig:grip-v1} and~\ref{fig:grip-v2}, which benchmark on MS MARCO v1 and MS MARCO v2, respectively. In Figure~\ref{fig:grip-v1} we observe that \textsc{GRIP} dominates HNSW on both SPLADE-CoCondenser and SPARSEUP, running $2.8$--$3.2\times$ and $2.0$--$2.1\times$ faster at recall@10 targets between 95\% and 98\% (though the gap narrows in the high-recall regime, falling to $1.5\times$ at 99\% on SPARSEUP). These results validate our sparsity-specific optimizations. Furthermore, we see that \textsc{GRIP} achieves a $1.4$--$1.7\times$ speedup over Seismic on SPLADE-CoCondenser across recall@10 targets from 90\% to 99\%. On SPARSEUP, \textsc{GRIP} is $1.2\times$ faster than Seismic at 92--95\% recall@10 and roughly on par at 97\%, while Seismic is faster in the extremely high-recall regime of 98\% and above. In Figure~\ref{fig:grip-v2}, the gap over Seismic widens at scale: on the 138M passages of MS MARCO v2, \textsc{GRIP} is $2.2$--$3.4\times$ faster than Seismic at recall@10 targets between 90\% and 98\%. This observation on the improvement of graph search over inverted indexes at larger scale is also consistent with the prior findings of \citet{bruch2025investigating}, who found that graphs exhibited better asymptotic scalability. However, at the time, the concrete efficiency of graph-based search still lagged behind inverted indexes like Seismic, so the point at which the scalability of graphs provided an advantage occurred at impractically large collection sizes. Our evaluation of \textsc{GRIP} shows that we can bend this scaling curve considerably by bringing ideas from inverted indexes into graph search. 

\begin{table}[t]              
    \centering                        
    \small                     
    \begin{tabular}{@{}lccc@{}}           
      \toprule                        
      & \multicolumn{2}{c}{MS MARCO v1 (8.8M)} & MS MARCO v2 (138M) \\                         
      \cmidrule(lr){2-3} \cmidrule(l){4-4} 
      Method & CoCondenser & SPARSEUP & CoCondenser \\        
      \midrule                        
      HNSW (efc 200)   & 19 min & 31 min & --- \\
      HNSW (efc 2000)  & 5.5 h  & 7.4 h  & --- \\
      Seismic          & 2 min & 3 min & 47 min \\
      \textsc{GRIP} & 47 min & 54 min & 7.3 h \\
      \bottomrule
    \end{tabular}
    \caption{Index construction time (30 threads). For Seismic, we report the build time for the index that gives its best latency at 95\% recall@10. We observe that Seismic has an order-of-magnitude faster build time than \textsc{GRIP}, underscoring that the inverted indexes still maintain a decisive advantage over graph-based indexes in terms of build time.}
    \label{tab:build-time}
  \end{table}

In Table~\ref{tab:build-time}, we report the index construction time of the index types we consider in our experiments. For the graph-based methods, \textsc{GRIP} and HNSW, we build a single index and generate the corresponding recall-latency tradeoff curve by varying search time parameters. For Seismic, we build many indexes to generate different latency-recall operating points, so we report the build time of the index that gives the best latency at 95\% recall@10. From Table~\ref{tab:build-time}, we see that \textsc{GRIP}'s Sparse PiPNN approach considerably reduces build time compared to HNSW with high efc, as previously noted. We also observe that Seismic maintains an order-of-magnitude faster build time than \textsc{GRIP} demonstrating that inverted indexes still maintain a considerable advantage over graph indexes in construction time. Nevertheless, the 7.3h construction time for \textsc{GRIP} at the 138M scale of MS MARCO v2 is still within practical constraints of search systems that require rebuilding indexes on a daily cadence. We do not report the build time of HNSW on MS MARCO v2 because an efc value of 200 would produce uncompetitive search quality and an efc of 2000 would require days to build. 

\begin{table}[t]
  \centering
  \small
  \begin{tabular}{@{}lccc@{}}
    \toprule
    & \multicolumn{2}{c}{MS MARCO v1 (8.8M)} & MS MARCO v2 (138M) \\
    \cmidrule(lr){2-3} \cmidrule(l){4-4}
    Method & CoCondenser & SPARSEUP & CoCondenser \\
    \midrule
    HNSW (efc 200)  & 5.5 & 7.8  & --- \\
    HNSW (efc 2000) & 6.3 & 8.5  & --- \\
    Seismic         & 6.5 & 16.1 & 73.0 \\
    \textsc{GRIP}          & 8.8     & 12.3    & 140 \\
    \bottomrule
  \end{tabular}
  \caption{Index size in GB. For
  Seismic, we report the size of the index that gives its best latency at 95\% recall@10.}
  \label{tab:index-size}
\end{table}

In Table~\ref{tab:index-size}, we report the index size for each of the methods we evaluate. On MS MARCO v1, we observe that the \textsc{GRIP} index consumes more space than Seismic with CoCondenser embeddings but less memory with SPARSEUP. We hypothesize that the lower space footprint on SPARSEUP stems from the fact that SPARSEUP embeddings are denser (more nonzero components) and have a larger vocabulary, leading Seismic's posting lists to grow faster than the graph. We leave further investigation of the impact of these parameters for future work. On MS MARCO v2 with CoCondenser, we note that \textsc{GRIP} uses almost double the space of Seismic primarily due to the fact that \textsc{GRIP} stores an additional copy of each vector in $R$-sketched form. Improving the space efficiency of \textsc{GRIP} is another promising direction for future work, following similar work on optimizing the space footprint of Seismic~\cite{nardini2026compact, bruch2026forward}.

\begin{table}[t]
  \centering
  \small
  \begin{tabular}{@{}llcccc@{}}
    \toprule
    & & \multicolumn{4}{c}{Recall@10} \\
    \cmidrule(l){3-6}
    Model & Search & 95\% & 97\% & 98\% & 99\% \\
    \midrule
    \multirow{3}{*}{CoCondenser}
      & Seeding only      & 2.39$\times$ & 2.59$\times$ & 2.55$\times$ & 2.34$\times$ \\
      & Routing only      & 1.37$\times$ & 1.38$\times$ & 1.34$\times$ & 1.22$\times$ \\
      & Seeding + routing & 2.97$\times$ & 3.36$\times$ & 3.32$\times$ & 3.07$\times$ \\
    \midrule
    \multirow{3}{*}{SparseUp}
      & Seeding only      & 1.99$\times$ & 1.95$\times$ & 1.96$\times$ & 1.95$\times$ \\
      & Routing only      & 1.16$\times$ & 1.10$\times$ & 1.12$\times$ & 1.16$\times$ \\
      & Seeding + routing & 2.15$\times$ & 2.08$\times$ & 2.04$\times$ & 2.17$\times$ \\
    \bottomrule
  \end{tabular}
  \caption{Ablation of \textsc{GRIP}'s search on MS MARCO v1 (8.8M). We report the speedup over standard beam search on the same Sparse PiPNN graph at matched recall@10 (single thread). Seeding uses $T{=}2$ (CoCondenser) and $T{=}4$
  (SparseUp) with $L{=}64$. Routing uses $R{=}80$ and $R{=}128$. We observe that term-based seeding provides the majority of the efficiency gains, but the two optimizations are complementary and combine to produce a larger speedup than either technique alone.}
  \label{tab:ablation}
\end{table}

Finally, in Table~\ref{tab:ablation}, we conduct an ablation study to isolate the speedup of term-based entry point seeding and approximate routing via sketching. We observe that term-based seeding provides larger gains than routing, but the two optimizations are complementary and, when combined, produce a larger speedup than either technique alone. 

\section{Conclusion}

Our core insight behind the design of GRIP is that we can infuse the strengths of inverted indexes into graph-based search. To that end, we introduced three optimizations that accelerate query processing and index construction, namely term-based entry point seeding, approximate routing via sparse sketching, and Sparse PiPNN, each of which is inspired by, or directly uses, inverted indexes. Together, these allow GRIP to outperform Seismic, a state-of-the-art inverted index, by 2.2–3.4× on the 138M-passage MS MARCO v2 collection with SPLADE-CoCondenser embeddings. These results challenge the recent consensus in the vector search community that inverted indexes are the better choice for learned sparse retrieval at realistic collection sizes. These gains do come at a cost: GRIP’s index is larger than Seismic’s and takes longer to build. Closing this gap, for instance through graph and vector compression, is an important direction for future work. Nevertheless, within LSR specifically, our results call into question the default choice of inverted indexes for deploying sparse retrieval, and may open the door to applying sparse embeddings in new, ambitious applications that demand more scalability. More broadly, we believe that our insights on combining the strengths of inverted indexes and graph-based indexes may have implications beyond learned sparse retrieval. Graph-based search for LSR was not inherently uncompetitive; it simply had not yet borrowed what made inverted indexes successful. We suspect the same is true in other data structures problems where one technique has been set aside in favor of another, and where the better answer may be to combine the strengths of both. Exploring the generality of our insights to other problems may also be a promising direction of future work.

\bibliographystyle{ACM-Reference-Format}
\bibliography{references}
\end{document}